\documentstyle[a4,epsfig,12pt]{article}
\begin{document}
\begin{center}
{\large\bf A model of anomalous production of strange baryons
in nuclear collisions}
\end{center}
\medskip
\begin{center}
{Roman Lietava${}^{a)d)}$, J\'an Pi\v{s}\'ut${}^{a)b)}$, Neva 
Pi\v{s}\'utov\'a${}^{a)}$ and Petr Z\'avada${}^{c)}$}
\end{center}
\medskip
\begin{center}
{\it a) Department of Physics MFF UK, Comenius University,
 SK-84215 Bratislava, Slovakia \\
     b) Laboratoire de Physique Corpusculaire, Universit\'e Blaise 
     Pascal, Clermont- Ferrand, F-63177 Aubi\`{e}re, Cedex, France\\ 
     c) Institute of Physics, Czech Academy of Science, Na Slovance 2, 
       CZ-18040 Praha 8, Czech Republic\\
     d) CERN,European Laboratory for Particle Physics,Geneva,Switzerland}
\end{center}
\vspace{1cm}

\abstract {We propose a simple model of production of strange baryons
and antibaryons in nuclear collisions at the CERN SPS. 
The model takes into account both the increase of strangeness
production in collisions of lighter ions and a possibility
of the formation of anomalous, strangeness rich matter in central
PbPb interactions. It is shown that
ratios like $<\Omega>:<\Xi>:<\Lambda>$ depend strongly on the
presence of anomalous matter and can be used to determine its
phenomenological parameters. In the model we assume that particle
composition of final state hadrons is essentially given by a
rapid recombination of quarks and antiquarks formed in  
tube-on-tube interactions of incoming nucleons.}

\vspace{1cm}

\vfill
\eject

\section{Introduction}
\label{intro}

Enhanced production of strange hadrons has been suggested
 \cite{RAF82,KOCH86} as a signature of Quark- Gluon Plasma (QGP)
formation in heavy ion collisions. Enhancement of strangeness production
 - with respect to pp interactions - has been observed \cite{NA35AA} in
pS, pAg, SS and SAg interactions, but it is in our opinion rather
unlikely that QGP has been formed in these collisions.
Strangeness enhancement observed in these interactions has been analysed
by Werner \cite{wer}, Sorge \cite{sor} and Tai An and Sa Ben-Hao
\cite{an} in models with collective string interactions and by Capella et
al. \cite{cap} in the dual parton model.

When analyzing the production of strange hadrons in Pb+Pb interactions
one has to disentangle that part of strangeness enhancement which is given
by the extrapolation of trends observed already in collisions of lighter
ions from genuine "anomalous" strangeness enhancement due to the production
of new form of matter in Pb+Pb collisions.

In Refs. \cite{wer,sor,an} the increase of strangeness due to interactions
of final state hadrons has been taken into account and found insufficient to
describe the observed enhancement without string-string effects. In
contradistinction to these authors we use phenomenological parametrization
of the production of $u\bar u$, $d\bar d$, $s\bar s$ pairs which will appear
in final state hadrons as valence quarks and antiquarks. This
parametrization is based on \cite{LIET97} and on the data analysis by
Wroblewski \cite{WRO85}. We have also taken into account a possible
production of anomalous strangeness rich matter in the spirit of Refs.
\cite{BLAIZ,KHARa,KHARb}

In the present paper we shall study  a simple model of the production of
strange baryons and antibaryons in proton- nucleus (pA)
 and nucleus- nucleus (AB)
interactions in the CERN SPS energy range. The model is based on the
assumption that yields of particles of different types are roughly given
by the recombination of quarks and antiquarks formed during the first stage
of the nuclear collision. According to this assumption the yields of
particles of different type are not strongly influenced by the stage of
interacting gas of hadrons.
 
  Models based on the  idea of recombination
   have been intensively studied
during past 15 years, see e.g. Refs.\cite{BIRO95}-\cite{BIZ83}.
An essential ingredient of any recombination model is the number of quarks
and anti- quarks (future valence quarks) just before the recombination.
 We shall estimate the number of
recombining quarks and antiquarks of the three flavours $u,d,s$ from
the analysis of data on production of strange and non- strange hadrons
in pA and AB interactions. Compilations of data can be found in Refs.
\cite{LIET97},\cite{JEON97}-\cite{MAL83}.

In what concerns the strange quarks and antiquarks we shall use the
parametrization of our recent work \cite{LIET97}, see also Refs.
\cite{WRO85,BIALK92}, the number of created $u\bar u$ and $d\bar d$ pairs
is described by a parametrization of the type used in Ref.\cite{LIET97}
and results of Wroblewski  \cite{WRO85}.

A rapid recombination differs from models based on the assumption
of a formation of a thermalized system in AB interactions, see e.g. Refs.
\cite{LET95}, although the free parameters in models permit to obtain
similar results. In particular the baryon chemical potential and
parameter of partial chemical thermalization of strange hadrons have their
counterparts in the ratio of quarks and antiquarks and the ratio of strange
to non-strange quarks before the thermalization.

     Assuming a rapid recombination implies  a rather short
pre-hadronic stage in heavy ion collisions, that means that hadrons
are formed within about 2fm/c, which is a time interval too short
to permit a transfer of quarks and antiquarks over distances in the
transverse plane comparable with the dimensions of the system - about
10fm.

An indirect support of a rather short pre- hadronic stage in nuclear
collisions  is provided by the phenomenological success
of models by Blaizot and
Ollitrault \cite{BLAIZ} and by Kharzeev, Lourenco, Nardi and Satz
\cite{KHARa,KHARb} in describing  $J/\Psi$ suppression,
observed by NA50 collaboration \cite{NA50a} in Pb+Pb collisions
at the CERN SPS. In models of Refs.\cite{BLAIZ,KHARa,KHARb} it is
assumed that  Quark- Gluon Plasma (QGP) is formed in a part of
impact parameter plane, separated from the rest where only
hadronic matter is present. Since the separation can hardly remain there
for a time of the order of 10 fm/c the existence of the boundary
indicates that the existence of
 QGP or, more generally,
of the form of matter which is strongly absorbing $J/\Psi $ is present
only for a short time of the order of 1-2 fm/c.

The present note is structured as follows. In the Sect.2 we shall
describe our parametrization of the numbers of produced pairs 
$u\bar u$, $d\bar d$ and
$s\bar s$ - in addition of valence quarks present in incoming
nuclei - in pA and AB interactions. 
In Sect.3 the rapid recombination scheme is considered.
A possible modification of the rapid
recombination scenario is briefly mentioned in Sect.4. 
In Sect.5 we discuss the modifications
of the number of created quark pairs due to the presence of "anomalous",
strangeness richer, matter. 
The last Sect.6
contains comments and conclusions.

\section{Parametrization of the production of $<u\bar u>$, $<d\bar d>$
and $<s\bar s>$ pairs in pA and AB interactions}
\label{para}

In our work \cite{LIET97} we have parametrized the production of
$<s\bar s>$ pairs in pA and AB interactions as
\begin{equation}
N_s^{coll}=<s\bar s>= <s\bar s>_{nn}\sum _{i,j}
(1-\beta_s)^{i-1}(1-\beta_s)^{j-1}
\label{eq1}
\end{equation}
where     the superscript "coll" indicates that $s\bar s$ pairs are
produced in nucleon- nucleon
 collisions and
$$<s\bar s>_{nn}(1-\beta_s)^{i-1}(1-\beta_s)^{j-1}
$$ 
is the number of
$<s\bar s>$ pairs produced in the nucleon- nucleon collision which
is i-th for the nucleon coming from the left and j-th for the nucleon
coming from the right.
 Values of  parameters $<s\bar s>_{nn}$ and $\beta_s$ are according 
 to data excluding
those for Pb+Pb collisions \cite{LIET97}
$$ \beta_s \approx 0.13,\qquad   <s\bar s>_{nn}\approx 0.63\pm 0.08           
$$
In Ref.\cite{WRO85} Wroblewski has analyzed the production of
future valence
quarks and antiquarks in hadron- hadron interactions,
 obtaining for the total number
of valence quarks and antiquarks
  in pp interactions at 205 GeV/c the result
$<N_{q\bar q}> = 7.4\pm 0.6$. Subtracting from that the number of 
$s\bar s$ pairs and dividing by two to get separately
$<u\bar u>_{nn}$ and $<d\bar d>_{nn}$ we obtain 
$$<d\bar d>_{nn}\approx <u\bar u>_{nn}\approx 3.4\pm 0.4$$ 
Making use of data
on $<h^->$ production in pA and AB interactions 
compiled by Gazdzicki and R\"{o}hrich \cite{GAZD96} 
and assuming that the number 
of quark- antiquark pairs is proportional to $<h^->$ we have obtained
$\beta_u =\beta_d \approx 0.4$. In this way we have
\begin{equation}
N_u^{coll}=<u\bar u>=<d\bar d>\approx 
<u\bar u>_{nn}\sum _{i,j}(1-\beta_u)^{i-1}(1-\beta_u)^{j-1}
\label{eq2}
\end{equation}
where
$$\beta_u=\beta_d\approx 0.4,\qquad 
<u\bar u>_{nn}=<d\bar d>_{nn}\approx 3.4\pm 0.4
$$

According to the scheme described above we can now write down
formulas for the average number of quark- antiquark pairs produced
in pA and AB interactions.

We shall start with pA collisions at a fixed value of the impact
parameter $b$. The number of $s\bar s$ pairs denoted as 
$N^{coll}_{s\bar s}(pA;b)$ is given as
$$
N^{coll}_{s\bar s}(pA;b)=<s\bar s>_{nn}\sum _{i=1}^{\mu}
      (1-\beta_s)^{i-1} =     
$$
\begin{equation}
=<s\bar s>_{nn}\frac {1-(1-\beta_s)^{\mu}}{\beta_s}
\label{eq3}
\end{equation}
where
$$
\mu = \sigma \rho 2L_A(b); \qquad L_A(b)=\sqrt {R_A^2 -b^2}
$$
Here $R_A$ is the radius of the nucleus A, $b$ is the impact parameter of
the collision, $\sigma $ is the inelastic nucleon- nucleon cross- section
and $\rho $ is the nuclear density.
Similarly for $u\bar u$ and $d\bar d$ pairs
$$
N^{coll}_{u\bar u}(pA;b)= N^{coll}_{d\bar d}(pA;b)=
<u\bar u>_{nn}\sum_{i=1}^{\mu} (1-\beta_u)^{i-1} =      
$$
\begin{equation}
=<u\bar u>_{nn}\frac {1-(1-\beta_u)^{\mu}}{\beta_u}
\label{eq4}
\end{equation}
where the notation is the same as in Eq.(3).

In addition to quark- antiquark pairs created in the interactions
there are also valence quarks of incoming nucleons. Averaging over
$d$ and $u$-quarks we have 1.5 of $d$-quark and 1.5 of $u$-quark for 
each participating nucleon. This gives
\begin{equation}
N^{val}_{u}= N^{val}_{d}=1.5(\mu +1); \qquad N^{val}_{s}
=0
\label{eq5}
\end{equation}
The total number of quarks and antiquarks taking part in the recombination
is given as the sum  of contributions from collisions and from
valence quarks
$$
N_u= N^{val}_{u}+N^{coll}_{u\bar u}
$$
\begin{equation}
N_d= N^{val}_{d}+N^{coll}_{d\bar d};\qquad N_s=N_{\bar s}=
 N^{coll}_{s\bar s}
\label{eq6}
\end{equation}

In the case of AB interactions we shall start with considering the number
of quarks and antiquarks due to the collisions of nucleons present in
two colliding tubes, each having cross section $\sigma$. The impact
 parameter of the collision is denoted as $\vec b$. In the transverse plain
  the position of the tube
 in nucleus A is specified by the vector $\vec s=\vec s_A$
   Within the nucleus B, the transverse position of the second tube
    is given by the
  vector $\vec s_B=\vec b-\vec s$. The average numbers of nucleons in
   tubes in A and B are respectively given as
$$
\nu \equiv \nu _A= \sigma \rho 2L_A(s); \qquad L_A(s)=\sqrt {R_A^2-s^2}
$$
\begin{equation}
\mu \equiv \mu _B=\sigma \rho 2L(\vec b,\vec s); \quad
2L_B(\vec b,\vec s)= \sqrt {R_B^2-b^2-s^2+2bs.cos(\theta )}
\label{eq7}
\end{equation}
Numbers of future valence quarks and antiquarks created in such a tube- on
-tube collision are given as
$$
N^{coll}_{s\bar s}(AB;b,s,\theta)=
<s\bar s>_{nn}\sum _{i=1}^{\nu } \sum _{j=1}^{\mu }
  (1-\beta _s)^{i-1}(1-\beta _s)^{j-1}=
$$
$$
=<s\bar s>_{nn}{\frac {1-(1-\beta_s)^{\nu}}{\beta _s}}
              {\frac {1-(1-\beta_s)^{\mu}}{\beta _s}}
$$
$$
N^{coll}_{u\bar u}(AB;b,s,\theta)=N^{coll}_{d\bar d}(AB;b,s,\theta)=
$$
$$
=<u\bar u>_{nn}\sum _{i=1}^{\nu } \sum _{j=1}^{\mu }
  (1-\beta _u)^{i-1}(1-\beta _u)^{j-1}=
$$
\begin{equation}
=<u\bar u>_{nn}{\frac {1-(1-\beta_u)^{\nu}}{\beta _u}}
              {\frac {1-(1-\beta_u)^{\mu}}{\beta _u}}
\label{eq8}
\end{equation}
The total number of future valence quarks and antiquarks produced  by
interaction within these tubes then becomes
$$
N_u(AB;b,s,\theta)=N_d(AB;b,s,\theta)= 1.5(\mu +\nu)+
                   N^{coll}_{u\bar u}(AB;b,s,\theta)
$$
$$
N_s(AB;b,s,\theta)=N_{\bar s}(AB;b,s,\theta)=
N^{coll}_{s\bar s}(AB;b,s.\theta)
$$
\begin{equation}
N_{\bar u}(AB;b,s,\theta)=N_{\bar d}(AB;b,s,\theta)= 
                   N^{coll}_{u\bar u}(AB;b,s,\theta)
\label{eq9}
\end{equation}

In other to obtain a qualitative feeling of ratios of different
flavours of quarks and antiquarks we plot in Table 1 values of
$N^{coll}_s$, $N^{coll}_u$, $N_u=N^{coll}_u$, and their ratios for
a set of values of $\mu $ and $\nu $.

\medskip
\begin{center}
{\bf Table 1.} Production of quarks and antiquarks in tube- on tube
collision; $\mu$ and $\nu$ are numbers of nucleons in the tubes.
\medskip

\begin{tabular}{|c|c|c|c|c|c|} \hline
$\mu$,$\nu$ & $N_s=N_s^{coll} $ & $N_u^{coll} $ &
 $N_u=N_u^{coll}+N_u^{val} $ & $N_s/N_u^{coll}$ & $N_s/N_u$  \\ \hline
 1,1   & 0.63  &  3.4  &  6.4  &  0.185  &  0.1   \\ \hline
 1,2   & 1.18  &  5.44 &  9.9  &  0.22   &  0.12  \\  \hline
 1,3   & 1.65  &  6.66 & 12.7  &  0.25   &  0.13  \\  \hline
 1,4   & 2.07  &  7.40 & 14.9  &  0.28   &  0.14  \\  \hline
 1,5   & 2.43  &  7.84 & 16.84 &  0.31   &  0.144  \\ \hline
 2,2   & 2.2   &  8.7  & 14.7  &  0.253  &  0.15   \\ \hline
 2,3   & 3.1   & 10.7  & 18.2  &  0.29   &  0.17   \\ \hline
 2,4   & 3.87  & 11.83 & 20.83 &  0.33   &  0.19   \\ \hline
 2,5   & 4.54  & 12.54 & 23.04 &  0.36   &  0.20   \\ \hline
 3,3   & 4.33  & 13.06 & 22.1  &  0.33   &  0.20   \\ \hline
 3,4   & 5.43  & 14.5  & 25.0  &  0.37   &  0.22   \\ \hline
 3,5   & 6.37  & 15.36 & 27.4  &  0.415  &  0.23   \\ \hline
 4,4   & 6.8   & 16.1  & 28.1  &  0.42   &  0.24   \\ \hline
 4,5   & 8.0   & 17.0  & 30.5  &  0.47   &  0.26   \\ \hline
 5,5   & 9.36  & 18.1  & 33.1  &  0.52   &  0.28   \\ \hline
 \end{tabular}
 \end{center}

As seen in the Table 1. the ratio $N_s/N_u$ increases by a factor
of 2.8 when going from the case of $\mu=1,\nu=1$ to that of
$\mu=5,\nu=5$. The increase of the production of strange to non- strange
hadrons will be, of course, smaller since in the nuclear collisions one
always integrates over region of nuclei overlap 
in the impact parameter plane and
the influence of the central region is suppressed by geometry.

The NA49 Collaboration has recently presented \cite{BOR97} results on 
$K^+/K^-$ and $\bar \Lambda/\Lambda $ ratios in the central Pb+Pb
collisions at 158 GeV per nucleon. The resulting numbers
$$
{{K^+}\over {K^-}}\approx 1.8; \qquad {{\bar \Lambda }\over {\Lambda}}
\approx 0.2
$$
 give a hint on whether the recombination models has a chance. Since
 $K^+$  consists of $s\bar u$,  $K^-$ of $\bar u s$,
 $\Lambda$ of $s,d,u$ and $\bar \Lambda$ of $\bar s,\bar d,\bar u$ we expect
 in a recombination model
\begin{equation}
{{K^-}\over {K^+}}\approx {{N_u}\over {N_{\bar u}}}.
                          {{N_{\bar s}}\over {N_s}} ; \quad
{{\bar \Lambda}\over {\Lambda}}\approx
                  {{N_{\bar s}}\over {{N_s}}}.
                  {{N_{\bar u}}\over {{N_u}}}.
                  {{N_{\bar d}}\over {{N_d}}}
\label{eq10}
\end{equation}
Table 1 shows that for most of combinations of $\mu$ and $\nu$ it
holds $N_u\approx 2N_{\bar u}$ and by assumption $N_s=N_{\bar s}$ so
both of ratios come out roughly correct. Note that in Table 1 we include
all the quarks and antiquarks just before recombination independently
of their rapidity. This may be adequate for the results of NA49 with
a large acceptance. 

\section{Production of hadrons via fast recombination}
\label{stran}

In this section we shall assume that the recombination is so fast that
 quarks and antiquarks produced in a given tube- on - tube collision
can recombine mutually  and what happens in a given tube- on tube system
is independent of what happens to systems produced by other tube- on-
tube collisions.

We shall use the recombination model suggested by Bir\'o and Zim\'anyi
 \cite{KOCH86,BIRO95,ZIM93,BIZ83}. According to this model the number of
 pions $N_{\pi}$, kaons $N_K$, $\phi$-mesons $N_{\phi}$, baryons $N_B$,
 $Y$-hyperons $N_Y$, $\Xi$-hyperons $N_{\Xi}$,  $\Omega$'s and corresponding
 antibaryons are given by the following relations:
$$ N_{\pi}= \alpha (N_u+N_d)(N_{\bar u}+N_{\bar d}), \qquad
   N_K=\alpha (N_u+N_d)N_{\bar s},\quad
$$
$$
   N_{\bar K}=\alpha (N_{\bar u}+N_{\bar d}),\qquad
   N_{\phi} = \alpha N_sN_{\bar s},
$$
$$ N_B=\beta {1\over {3!}}(N_u+N_d)^3,\qquad
   N_Y=\beta {1\over {2!}}(N_u+N_d)^2 N_s,
$$
\begin{equation}
   N_{\Xi}=\beta {1\over {2!}} N_s^2(N_u+N_d),\qquad
   N_{\Omega}=\beta {1\over {3!}}N_s^3.
\label{eq11}
\end{equation}
The constants $\alpha $ and $\beta $ are obtained from the consistency
conditions requiring that the number of quarks and antiquarks is equal
to the corresponding number of valence quarks and antiquarks in hadrons
formed by recombination.
In this way one finds 
 \cite{KOCH86,BIRO95,ZIM93,BIZ83}
\begin{equation}
\alpha= \frac {Q+\bar Q}{Q^2+Q\bar Q+\bar Q^2}; \quad
\beta = \frac {2}{Q^2+Q\bar Q +\bar Q^2}
\label{eq12}
\end{equation}
where
$$
Q=N_u+N_d+N_s; \qquad  \bar Q = N_{\bar u}+N_{\bar d}+N_{\bar s}
$$

The yield of a certain particle is calculated in the following way.
Numbers of quarks and antiquarks of all flavours in a tube- on- tube
collision at given $(b,s,\theta)$ are obtained via Eqs.(8,9). This is
inserted into Eqs.(11,12). In these equations we have used a short hand
notation like $N_{\pi}$. The full notation should be $N_{\pi}(b,s,\theta)$.
The total yield of $Y$-hyperons is then obtained from the expression
\begin{equation}
N_Y(AB;b)={1\over {\sigma}}\int_0^{R_A}sds
\int _0^{2\pi}d\theta N_Y(AB;b,s,\theta)
\label{eq13}
\end{equation}
We have calculated the yields $N_Y(AB,b)$ by two independent numerical
methods which gave very similar results.
The expression $N_Y(AB;b) $ corresponds to the sum of hyperons $\Lambda$,
$\Sigma ^-$, $\Sigma ^0$ and $\Sigma ^-$. Taking into account the decay
$\Sigma ^0 \to \Lambda +\gamma$ and the decay $\Xi ^0 \to \Lambda \pi^0$
 we have
 $$  <\Lambda >={1\over 2}N_Y+{1\over 2}N_{\Xi}
 $$
Although we shall make in this paper no attempts at comparison with the
data, let us point out that previous relationship corresponds to the
situation in NA49 data analysis, but not to the one in WA97.
 Similarly the experimentally observed number of $\Xi^-$ is given as
 $$ <\Xi^->={1\over 2} N_{\Xi}
 $$
 The results for Pb+Pb and S+S are presented in Tables 2 and 3. 

\medskip
\begin{center}
{\bf Table 2.} 
Yields of strange baryons and antibaryons
 in Pb+Pb collisions
as a function of the impact parameter $b$ in the
model of rapid recombination.
\medskip

\begin{tabular}{|c|c|c|c|c|c|c|} \hline
      &        &       &      &       &       &          \\
 b &$N_Y$ & $N_{\Xi}$ & $N_{\Omega}$ & $N_{\bar Y}$ & $N_{\bar \Xi}$ &
 $N_{\bar \Omega} $   \\ \hline
 0.0  & 141.4  & 18.62 & 0.84 & 43.6  & 10.27  & 0.84    \\ \hline
 1.0  & 136.1  & 18.02 & 0.82 & 41.64 &  9.91  & 0.82    \\ \hline
 2.0  & 125.83 & 16.68 & 0.75 & 38.1  &  9.14  & 0.75    \\ \hline
 3.0  & 113.1  & 14.9  & 0.67 & 33.9  &  8.15  & 0.67    \\ \hline
 4.0  &  98.9  & 12.92 & 0.58 & 29.6  &  7.05  & 0.58    \\ \hline
 5.0  &  84.13 & 10.83 & 0.47 & 25.1  &  5.92  & 0.47    \\ \hline
 6.0  &  69.36 &  8.74 & 0.38 & 20.8  &  4.79  & 0.38    \\ \hline
 7.0  &  55.1  &  6.76 & 0.28 & 16.63 &  3.72  & 0.28    \\ \hline
 8.0  &  41.8  &  4.94 & 0.20 & 12.74 &  2.74  & 0.20    \\ \hline
 9.0  &  29.8  &  3.37 & 0.13 &  9.22 &  1.88  & 0.13    \\ \hline
10.0  & 19.4   &  2.07 & 0.075&  6.11 &  1.16  & 0.075   \\ \hline
11.0  & 11.0   &  1.09 & 0.036&  3.54 &  0.62  & 0.036   \\ \hline
12.0  &  4.93  &  0.43 & 0.013&  1.6  &  0.25  & 0.013   \\ \hline
\end{tabular}               
\end{center}

\begin{center}
{\bf Table 3.} 
Yields of strange baryons and antibaryons
 in S+S collisions
as a function of the impact parameter $b$ in the
model of rapid recombination.
\medskip

\begin{tabular}{|c|c|c|c|c|c|c|} \hline
      &        &         &        &        &         &       \\
 b &$N_Y$ & $N_{\Xi}$ & $N_{\Omega}$ & $N_{\bar \Lambda}$ &$N_{\bar \Xi}$ &
 $N_{\bar \Omega} $   \\ \hline      
 0.0  &  16.44  &  1.41  &  0.041  &  5.71  &  0.83  & 0.041  \\ \hline
 1.0  &  14.85  &  1.28  &  0.038  &  5.12  &  0.75  & 0.038  \\ \hline
 2.0  &  11.97  &  1.02  &  0.030  &  4.08  &  0.60  & 0.030  \\ \hline
 3.0  &   8.72  &  0.72  &  0.020  &  2.93  &  0.42  & 0.020  \\ \hline
 4.0  &   5.59  &  0.44  &  0.012  &  1.85  &  0.26  & 0.012  \\ \hline
 5.0  &   2.94  &  0.21  &  0.0053 &  0.96  &  0.12  & 0.0053 \\ \hline
 6.0  &   1.07  &  0.068 &  0.0015 &  0.33  &  0.038 & 0.0015 \\ \hline
\end{tabular}
\end{center}

The translation from the impact parameter $b$ to the number of nucleon-
nucleon collisions $N_c(b)$ and to the number of participating (wounded)
nucleons $N_p(b)$ is given by the standard relations
$$
N_p(b)={1\over {\sigma}}\int _0^{R_A}sds\int _0^{2\pi}d\theta
\Theta (R_B^2-b^2-s^2+2bs.cos\theta)
$$
$$
   \left(\rho_A\sigma 2L_A(s)[1-e^{-\rho_B\sigma 2L_B(b,s,\theta)}]+
        \rho_B\sigma 2L_B(b,s,\theta)[1-e^{-\rho_A\sigma 2L_A(s)}]\right)
$$
$$
N_c(b)={1\over {\sigma}}\int_0^{R_A}sds\int_0^{2\pi}d\theta
       \rho_A\sigma 2L_A(s)\rho_B\sigma 2L_B(b,s,\theta)
$$
where
$$
L_A(s)=\sqrt{R_A^2-s^2},\quad L_B(b,s,\theta)=\sqrt{R_B^2-b^2-s^2
        +2bs.cos(\theta)}
$$
The average values of strange baryons are calculated by using the
relations
$$ <\Lambda>={1\over 2} (N_Y+N_{\Xi}),\quad <\Xi^->={1\over 2}N_{\Xi},
\quad <\Omega>=N_{\Omega}
$$
and similarly for antibaryons. In this way we obtain results presented
in Tables 4 and 5.

\begin{center}
{\bf Table 4.} Yields of strange baryons and antibaryons
 in Pb+Pb collisions
as a function of the impact parameter $b$ in the
model of rapid recombination within the tubes.
\medskip

\begin{tabular}{|c|c|c|c|c|c|} \hline
      &        &     &        &       &                 \\
 b &$<\Lambda>$ & $<\Xi^->$ & $<\Omega>=<\bar \Omega>$
  & $<\bar \Lambda>$ & $<\bar \Xi^+>$    \\  \hline
 0.0  & 80.01  &  9.31 & 0.84 & 26.95 &  5.13     \\ \hline
 1.0  & 77.07  &  9.01 & 0.82 & 25.78 &  4.95     \\ \hline
 2.0  & 71.26  &  8.34 & 0.75 & 23.97 &  4.57     \\ \hline
 3.0  & 63.99  &  7.46 & 0.67 & 21.04 &  4.07    \\ \hline
 4.0  & 55.90  &  6.46 & 0.58 & 18.3  &  3.53    \\ \hline
 5.0  & 47.48  &  5.42 & 0.47 & 15.53 &  2.96    \\ \hline
 6.0  & 39.05  &  4.37 & 0.38 & 12.79 &  2.40     \\ \hline
 7.0  & 30.92  &  3.38 & 0.28 & 10.18 &  1.86    \\ \hline
 8.0  & 23.23  &  2.97 & 0.20 &  7.74 &  1.37    \\ \hline
 9.0  & 16.57  &  1.68 & 0.129&  5.55 &  0.94     \\ \hline
10.0  & 10.73  &  1.033& 0.075&  3.64 &  0.582    \\ \hline
11.0  &  6.054 &  0.542& 0.036&  2.08 &  0.308    \\ \hline
12.0  &  2.68  &  0.216& 0.013&  0.926&  0.124   \\ \hline
\end{tabular}               
\end{center}

In Fig.1 we present the dependence of the production of
of $<\Lambda>+<\bar \Lambda>$, $<\Xi^->+<\bar \Xi^+>$ and
$<\Omega>+<\bar \Omega>$
 on the number of nucleon- nucleon collisions. All yields
are normalized to 1 at the impact parameter value of b=10,
corresponding to the number of nucleon- nucleon collisions
$N_c(b=10)$=104.5. In the next section these results will be compared
 with the situation when anomalous and strangeness richer matter is
 present.
In order to permit a comparison with earlier work we present in Table 5.
the yields of strange baryons and antibaryons in S+S collisions.

\begin{center}
{\bf Table 5.} Yields of strange baryons and antibaryons in S+S
interactions as a function of the impact parameter $b$
\medskip

\begin{tabular}{|c|c|c|c|c|c|c|} \hline
      &        &         &        &               &       \\
 b &$<\Lambda>$ & $<\Xi^->$ & $<\Omega>=<\bar \Omega> $ &
  $<\bar \Lambda>$ & $<\bar \Xi^+>$       \\ \hline      
 0.0  &   8.93  &  0.71  &  0.041  &  3.27  &  0.42      \\ \hline
 1.0  &   8.07  &  0.64  &  0.038  &  2.94  &  0.38      \\ \hline
 2.0  &   6.50  &  0.51  &  0.030  &  2.34  &  0.30      \\ \hline
 3.0  &   4.72  &  0.36  &  0.020  &  1.68  &  0.21      \\ \hline
 4.0  &   3.01  &  0.22  &  0.0118 &  1.055 &  0.128      \\ \hline
 5.0  &   1.58  &  0.107 &  0.0053 &  0.54  &  0.062     \\ \hline
 6.0  &   0.567 &  0.034 &  0.0015 &  0.185 &  0.019     \\ \hline
\end{tabular}
\end{center}

Results for central S+S interactions can be compared with the data
\cite{NA35AA} and with the calculations performed within the ALCOR
model by Bir\'o, L\'evai and Zim\'anyi \cite{BIRO95}. Bir\'o et al.
obtain $<\Lambda>$=10.37, the data give $<\Lambda>=9.4\pm 1.0$ and our
value in Table 5. is $<\Lambda>$=8.93. For $<\Xi^->$ Bir\'o et al.
find 1.15 whereas our value is 0.73. The largest discrepancy between
Ref.\cite{BIRO95} and our results appears in $\Omega $ production
where the authors of Ref.\cite{BIRO95} find $<\Omega>$=0.14, whereas our
value is 0.041.

In the case of central Pb+Pb interactions Bir\'o et al. have obtained
$<\Lambda>$=82.4 and $<\Lambda>$=111.3 in the two versions of their model.
Our result in Table 5 is $<\Lambda>$=80.0.

\medskip
\section{Production of hadrons by a slow recombination}
\label{slow}

In the  the previous section we have considered a model in which both the
formation of future valence quarks and antiquarks and their recombination
to hadrons takes place within tubes of the cross- section equal to the
inelastic nucleon- nucleon cross- section $\sigma $. What happens in one
tube is in this model completely independent of what happens in another
tube. In order to have a qualitative feeling of the effects of this 
assumption we shall discuss in this section a model in which all quarks
and antiquarks formed in an A+B interaction at a given value of
the impact parameter $b$ can recombine with each other. The calculations
proceed as above but the order is reversed. By using Eqs.(8,9)
we compute numbers of future valence quarks produced in individual
tube- on- tube interactions. In the next step we calculate total numbers
of quarks and antiquarks formed in the A+B collision at given value of
the impact parameter $b$ according to equations like
\begin{equation}
N^T_u(AB;b)= {1\over {\sigma}}\int _0^{R_A} sds \int _0^{2\pi}d\theta 
             N_u(AB;b,s,\theta )
\label{eq17}
\end{equation}
The obtained total numbers of quarks and antiquarks are then recombined
via the Bir\'o - Zim\'anyi scheme as given by Eqs.(11) and (12).
The results are presented in Table 6. 
They are very close to those shown in Table 2. This fact is easyly
understood for the case of $Y$ hyperons.
Looking in Table 1. we see that with very good
approximation 
$N_u\approx 2N_{\bar u} >> N_s $ and $N_u=N_d$. 
Then parameter $\beta$ in Eq. (12) is $\beta \propto 1/N_u^2$
and $N_Y\propto N_s$
Now the number of hyperons in rapid recombination we get
by calculating number of $N_Y$ in each row and summing
over all rows.
$$ N_Y^{rapid}=\sum_{rows} N_Y^{row}=const.\sum_{rows} N_s^{row}
=const. N_s=N_Y^{slow}$$.
Similar arguments can be done for $\Xi$ and $\Omega$ hyperons.

\medskip
\begin{center}
{\bf Table 6.} Yields of strange baryons and antibaryons
 in Pb+Pb collisions
as a function of the impact parameter $b$ in the
model of slow recombination. 
\medskip

\begin{tabular}{|c|c|c|c|c|c|c|} \hline
      &        &       &      &       &       &          \\
 b & $N_Y$& $N_{\Xi}$& $N_{\Omega }$ & $N_{\bar Y}$ &$N_{\bar \Xi}$ &
 $N_{\bar \Omega }$   \\ \hline
 0.0  & 142.3  & 18.14 & 0.77 & 44.34 & 10.13  & 0.77    \\ \hline
 1.0  & 136.9  & 17.63 & 0.76 & 42.18 &  9.78  & 0.76    \\ \hline
 2.0  & 126.6  & 16.37 & 0.71 & 38.46 &  9.02  & 0.71    \\ \hline
 3.0  & 113.8  & 14.67 & 0.63 & 34.19 &  8.04  & 0.63    \\ \hline
 4.0  &  99.5  & 12.72 & 0.54 & 29.71 &  6.95  & 0.54    \\ \hline
 5.0  &  84.7  & 10.67 & 0.45 & 25.22 &  5.82  & 0.45    \\ \hline
 6.0  &  69.9  &  8.63 & 0.36 & 20.83 &  4.71  & 0.36    \\ \hline
 7.0  &  55.5  &  6.67 & 0.27 & 16.64 &  3.65  & 0.27    \\ \hline
 8.0  &  42.1  &  4.88 & 0.19 & 12.72 &  2.69  & 0.19    \\ \hline
 9.0  &  30.00 &  3.33 & 0.12 &  9.19 &  1.84  & 0.12    \\ \hline
10.0  & 19.5   &  2.05 & 0.07 &  6.08 &  1.14  & 0.07    \\ \hline
11.0  & 11.1   &  1.08 & 0.03 &  3.51 &  0.60  & 0.03    \\ \hline
12.0  &  4.97  &  0.43 & 0.01 &  1.59 &  0.24  & 0.01    \\ \hline
\end{tabular}
\end{center}

\section{ Possible presence of anomalous matter and thresholds in 
  strange baryon and antibaryon production}
\label{anomaly}

Data on the multiplicity of negative secondary hadrons and
on the total transverse energy  in Pb+Pb
 interactions \cite{NA49a} do not indicate a presence
of some thresholds connected with the formation of 
a new "anomalous" form of matter. It rather seems that
 the multiplicity of
 secondary hadrons in Pb+Pb
and transverse energy  can be obtained as the extrapolation
 of results obtained in collisions of lighter ions.

 On the other
 hand recent data of the WA97 Collaboration  indicate that the
 production of strange baryons
 within the acceptance region of the experiment
  is increased \cite{WA97a,KRAL97,WA98}.

 The WA97 experiment takes data only in a small part of the
 total phase- space. The accepted events cover a region near the central
 rapidity in the c.m.s. and transverse momenta of baryons above 0.6 GeV/c.
 In order to disentangle the extrapolated strangeness content one would need
 to use data from lighter ion collisions
 in the same experiment to determine the values of
 parameters $\beta_u$, $\beta_s$, $<u\bar u>_{nn}$ and $<s\bar s>_{nn}$
 corresponding to the experimental acceptance.

 Presence of a threshold in the production of strange baryons
 together with approximately no increase in the total multiplicity
  lead to the assumption that the total number of quark-
 antiquark pairs in tube- on- tube collisions in Pb+Pb interactions
 is  approximately the same as calculated by the
 formulas given above, but starting
 with tube- on- tube collisions which satisfy a certain criticality
 condition the matter is in some sense "melted" and a part of
 $u\bar u$ and $d\bar d$ pairs is transformed to $s\bar s$ pairs.

 In order to permit a comparison with the description of
the anomalous  $J/\Psi$  suppression in Pb+Pb in models of
Blaizot and Ollitrault \cite{BLAIZ} and of Kharzeev et al. \cite{KHARa}
we shall use the criticality condition of Ref.\cite{KHARa}.

In their description of $J/\Psi$ suppression Kharzeev et al.
\cite{KHARa,KHARb} assume that QGP is formed only in a limited region
of the transverse plane. Taking that view we shall introduce the
parameter
\begin{equation}
\kappa (b,s,\theta)= \frac {\sigma_{nn}\rho_A 2L_A(s).\sigma_{nn}\rho_B
                            2L_B(b,s,\theta)}
     {\sigma_{nn}\rho_A 2L_A(s)+\sigma_{nn}\rho_B 2L_B(b,s,\theta)}
\label{eq14}
\end{equation}
The parameter $\kappa(b,s,\theta)$ is roughly proportional  in the
"tube- on -tube picture" to the ratio 
of the number of nucleon- nucleon collisions to the         
longitudinal dimension of the system formed by the two colliding tubes.
It is further assumed that for 
\begin{equation}
\kappa(b,s,\theta) \ge \kappa_{crit}
\label{eq15}
\end{equation}
QGP is formed, whereas in the opposite case the system remains in the
normal state. 

  In Ref.\cite{KHARa}
the authors introduce two values of $\kappa_{crit}$. Above 
$\kappa_{crit}\approx $2.3 QGP is formed and the $\chi$ meson 
responsible for about 40\% of $J/\Psi$ production is dissolved.
For $\kappa_{crit}$ above 2.9 also $J/\Psi$ is completely dissolved.

In the present work  we shall use a single threshold, corresponding to
the onset of a "new" or "anomalous" form of matter. According to
Refs.\cite{KHARa,KHARb} we expect this threshold at about
 $\kappa_{crit}$=2.3.
 We assume that
above this threshold the matter goes into a new form 
characterized by a higher value of the strangeness abundance, that
means that above $\kappa _{crit}$ a part of $u\bar u$ and $d\bar d$
pairs is transformed to $s\bar s$ ones.

To simulate this effect, we go back to the Eqs.(8,9) and for $\kappa \ge \kappa_{crit}$
make the replacement
$$
N^{coll}_{s\bar s} \to N^{coll}_{s\bar s}
       [1+(\xi -1)\Theta (\kappa-\kappa_{crit})]
$$
$$ N^{coll}_{u\bar u} \to N^{coll}_{u\bar u} -
            0.5(\xi -1)N^{coll}_{s\bar s}\Theta (\kappa-\kappa_{crit})
$$ 
\begin{equation}
N^{coll}_{d\bar d} \to N^{coll}_{s\bar s}-
            0.5(\xi -1)N^{coll}_{s\bar s}\Theta (\kappa-\kappa_{crit})
\label{eq16}
\end{equation}
and then continue as within the scheme of the rapid recombination model 
in Sect.3. In this way we obtain for the two cases considered the results
presented in Tables 7 and 8. In order to see the effects caused by the
presence of the anomalous matter these results should be compared with
those given in Table 2.

\begin{center}
{\bf Table 7.} Yields of strange baryons and antibaryons
 in Pb+Pb collisions
as a function of the impact parameter $b$ within the
model of rapid recombination.
 Anomalous matter present: $\xi=2.0$, $\kappa_c$ =2.1.
\medskip

\begin{tabular}{|c|c|c|c|c|c|c|} \hline
      &        &       &      &       &       &         \\
 b &$N_Y$ & $N_{\Xi}$ & $N_{\Omega}$ & $N_{\bar Y}$ &$N_{\bar \Xi}$ &
 $N_{\bar \Omega} $   \\ \hline
 0.0  & 186.8  & 52.6  & 5.59 & 45.95 & 24.96  & 5.59    \\ \hline
 1.0  & 179.8  & 50.7  & 5.40 & 43.85 & 24.08  & 5.40    \\ \hline
 2.0  & 166.6  & 47.2  & 5.00 & 40.17 & 22.30  & 5.00    \\ \hline
 3.0  & 149.6  & 42.00 & 4.41 & 35.88 & 19.87  & 4.41    \\ \hline
 4.0  & 130.3  & 35.9  & 3.72 & 31.35 & 17.04  & 3.72    \\ \hline
 5.0  & 110.1  & 29.4  & 2.98 & 26.77 & 14.05  & 2.98    \\ \hline
 6.0  &  89.6  & 22.8  & 2.24 & 22.23 & 11.04  & 2.24    \\ \hline
 7.0  &  69.5  & 16.5  & 1.55 & 17.82 &  8.12  & 1.55    \\ \hline
 8.0  &  50.7  & 10.8  & 0.94 & 13.6  &  5.42  & 0.94    \\ \hline
 9.0  &  33.9  &  5.9  & 0.44 &  9.68 &  3.07  & 0.44   \\ \hline
10.0  & 19.4   &  2.07 & 0.08 &  6.11 &  1.16  & 0.08    \\ \hline
11.0  & 11.0   &  1.09 & 0.04 &  3.54 &  0.62  & 0.04    \\ \hline
12.0  &  4.93  &  0.43 & 0.01 &  1.6  &  0.25  & 0.01    \\ \hline
\end{tabular}
\end{center}

\medskip
\begin{center}
{\bf Table 8.} Yields of strange baryons and antibaryons
 in Pb+Pb collisions
as a function of the impact parameter $b$ within the
model of rapid recombination. Anomalous matter present: $\xi$=2.0;
 $\kappa_c$=2.5.
\medskip

\begin{tabular}{|c|c|c|c|c|c|c|} \hline
      &        &       &      &       &       &         \\
 b &$N_Y$ & $N_{\Xi}$ & $N_{\Omega}$ & $N_{\bar Y}$&$N_{\bar \Xi}$ &
 $N_{\bar \Omega} $   \\ \hline
 0.0  & 167.7  & 40.04 & 3.99 & 44.13 & 19.17  & 3.99    \\ \hline
 1.0  & 161.8  & 38.89 & 3.88 & 42.16 & 18.59  & 3.88    \\ \hline
 2.0  & 149.4  & 35.71 & 3.53 & 38.59 & 17.06  & 3.53    \\ \hline
 3.0  & 133.4  & 31.20 & 3.03 & 34.44 & 14.96  & 3.03    \\ \hline
 4.0  & 115.1  & 25.77 & 2.42 & 30.03 & 12.46  & 2.42    \\ \hline
 5.0  &  95.91 & 19.98 & 1.78 & 25.55 &  9.80  & 1.78    \\ \hline
 6.0  &  76.30 & 14.02 & 1.11 & 21.20 &  7.06  & 1.11    \\ \hline
 7.0  &  57.21 &  8.33 & 0.50 & 16.75 &  4.41  & 0.50    \\ \hline
 8.0  &  41.8  &  4.94 & 0.2  & 12.74 &  2.74  & 0.2     \\ \hline
 9.0  &  29.8  &  3.37 & 0.13 &  9.22 &  1.88  & 0.13    \\ \hline
10.0  & 19.4   &  2.07 & 0.08 &  6.11 &  1.16  & 0.08    \\ \hline
11.0  & 11.0   &  1.09 & 0.04 &  3.54 &  0.62  & 0.04    \\ \hline
12.0  &  4.93  &  0.43 & 0.01 &  1.6  &  0.25  & 0.01    \\ \hline
\end{tabular}
\end{center}

 Proceeding as below the Table 2, we obtain
 the yields of $<\Lambda>+<\bar \Lambda>$, $<\Xi_->+<\bar \Xi^+>$ and
 $<\Omega>+<\bar \Omega>$ normalized to 1 at $b=10$ that means
 $N_c$=104,5 as presented in Figs.2 and 3.

The rapid increase of the yields of $<\Lambda>+<\bar \Lambda>$,
$<\Xi^->+<\bar \Xi^+>$ and in particular of $<\Omega>+<\bar \Omega>$
seen in Figs.2 and 3. in comparison with Fig.1 shows
that the presence  of anomalous matter increases the yields of
strange hyperons significantly. The two parameters $\kappa_c$ and
$\xi $ regulate the position of the onset of the increase as a function
of the impact parameter or alternatively as a function on the number
of nucleon- nucleon collisions or the number of participating nucleons.
These two parameters should be determined by the comparison of model
predictions with the data. We shall not attempt to make this comparison
here.

Let us remark that a similar calculation was done also for slow
recombination scenario.
Comparison of the results obtained in both recombination schemes shows
greater differences between them for the case of the anomalous enhanced
strangeness production.
This behaviour is natural since the difference in final hyperon
multiplicities between slow
and fast recombination depends mainly on the ratio of strange
to nonstrange quarks.
We shall not discuss this issue in more
detail since we consider the rapid recombination scheme as a more
realistic model.

\section{Summary, comments and conclusions}
\label{comments}
 
We have presented here a recombination model of hadron formation in
nuclear collisions, which permits to calculate hadron yields as a function
of the impact parameter. The model is based on the phenomenological
parametrization of the number of quarks and antiquarks just before the
hadronization. A few parameters are determined by comparison with data
on hadron production in interactions of lighter ions and extrapolated to
the case of Pb+Pb collisions. The second ingredient is the Bir\'o -
Zim\'anyi recombination scheme.

The model contains several simplifications. In its present form it does
not contain fluctuations in the number of produced quarks and antiquarks
which may be important in estimating yields of multistrange baryons
in pA interactions and in collisions of lighter ions. For this reason we
have not normalized the strange baryon yields to pA interactions.

The model  does not analyse rapidity and $p_T$ distributions of quarks
and antiquarks before the recombination and therefore gives only predictions
for the total numbers of final state hadrons. Because of that, when
comparing model predictions with the data in a small part of the phase
space, one should rather determine the values of input parameters by data
on hadron production in the corresponding acceptance region.

The model  neglects modifications of the chemical composition of final
state hadrons due to interactions in the hadronic stage of the nuclear
collisions. The changes in the chemical composition in the hadronic phase
are known to be slow, and we do not expect that
 they will modify substantially
the yields of strange baryons.

On the other hand, the model includes a phenomenological description of
the influence of the presence of anomalous, strangeness rich, matter.
The parametrization of the effects due to the anomalous matter (perhaps
QGP) is similar to models used by Blaizot and Ollitrault, and Kharzeev
and Satz to describe the anomalous $J/\Psi$ suppression in Pb+Pb
interactions. The analysis of data on strange baryon production within this
model can thus contribute to the understanding 
 of the anomalous $J/\Psi$ suppression as observed by the NA 50
collaboration.

We have made here no attempt to compare our results with the data
of the WA97 collaboration covering the midrapidity region. This 
would require a specification of what fraction of valence quarks
participate in the recombination to strange baryons in this region.
We shall return to this question in the near future. 

\medskip
{\large \bf Acknowledgements} We are indebted to W.Geist, R.C.Hwa,
I.Kr\'alik, M.Moj\v{z}i\v{s}, E.Quercigh, L.\v{S}\'andor and other members
of the WA97 Collaboration for numerous useful discussions. One of the
authors (J.P.) would like to thank G.Roche and B.Michel for the hospitality
at the Laboratoire de Physique Corpusculaire at the Blaise Pascal
 University at Clermont- Ferrand, where a part of this work has been done.
 This work was supported in part by the US- Slovak Science and
 Technology Joint Fund under grant No. 003-95 and by the Slovak Grant
 Agency under grants No. V2F13-G and V2F18-G.

\medskip
{\large \bf Figure Captions}

\medskip
{\bf Fig.1.}

Yields of $<\Lambda>+<\bar \Lambda>$, $<\Xi^->+<\bar \Xi^+>$ and
$<\Omega>+<\bar \Omega>$ as a function of the number of nucleon-
nucleon collisions $N_c$ in the case when no anomalous matter is present.
All yields normalized to 1 at $b=10$ that means at $N_c$=104,5.

\medskip
{\bf Fig.2.}

Yields of $<\Lambda>+<\bar \Lambda>$, $<\Xi^->+<\bar \Xi^+>$ and
$<\Omega>+<\bar \Omega>$ as a function of the number of nucleon-
nucleon collisions $N_c$ in the case when  anomalous matter
characterized by parameters $\kappa_c=2.5,\xi=2.0$ is present.
All yields normalized to 1 at $b=10$, that means at $N_c$=104.5.

\medskip
{\bf Fig.3.}

Yields of $<\Lambda>+<\bar \Lambda>$, $<\Xi^->+<\bar \Xi^+>$ and
$<\Omega>+<\bar \Omega>$ as a function of the number of nucleon-
nucleon collisions $N_c$ in the case when  anomalous matter
characterized by parameters $\kappa_c=2.0,\xi=2.0$ is present.
All yields normalized to 1 at $b=10$, that means at $N_c$=104.5.

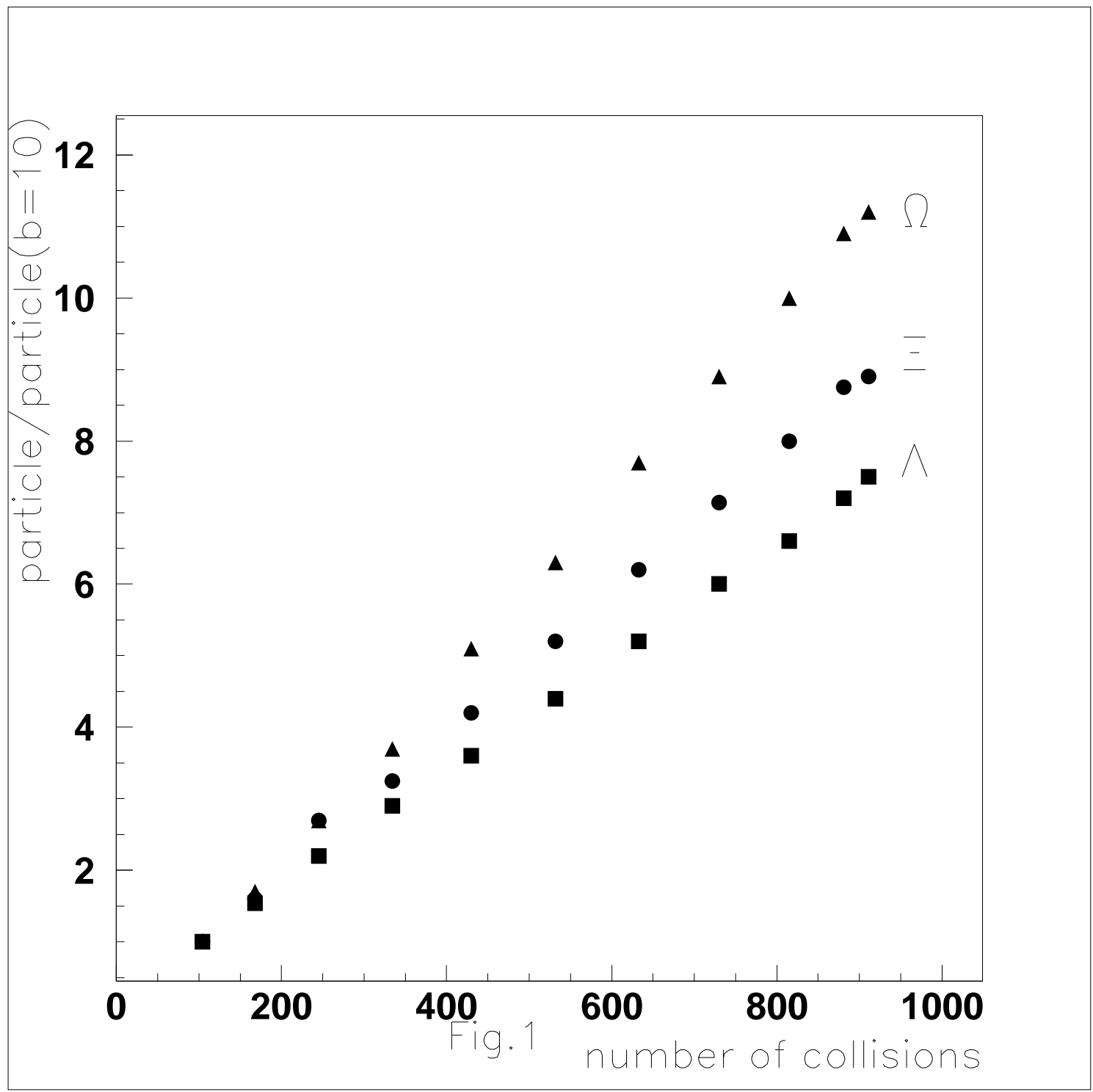
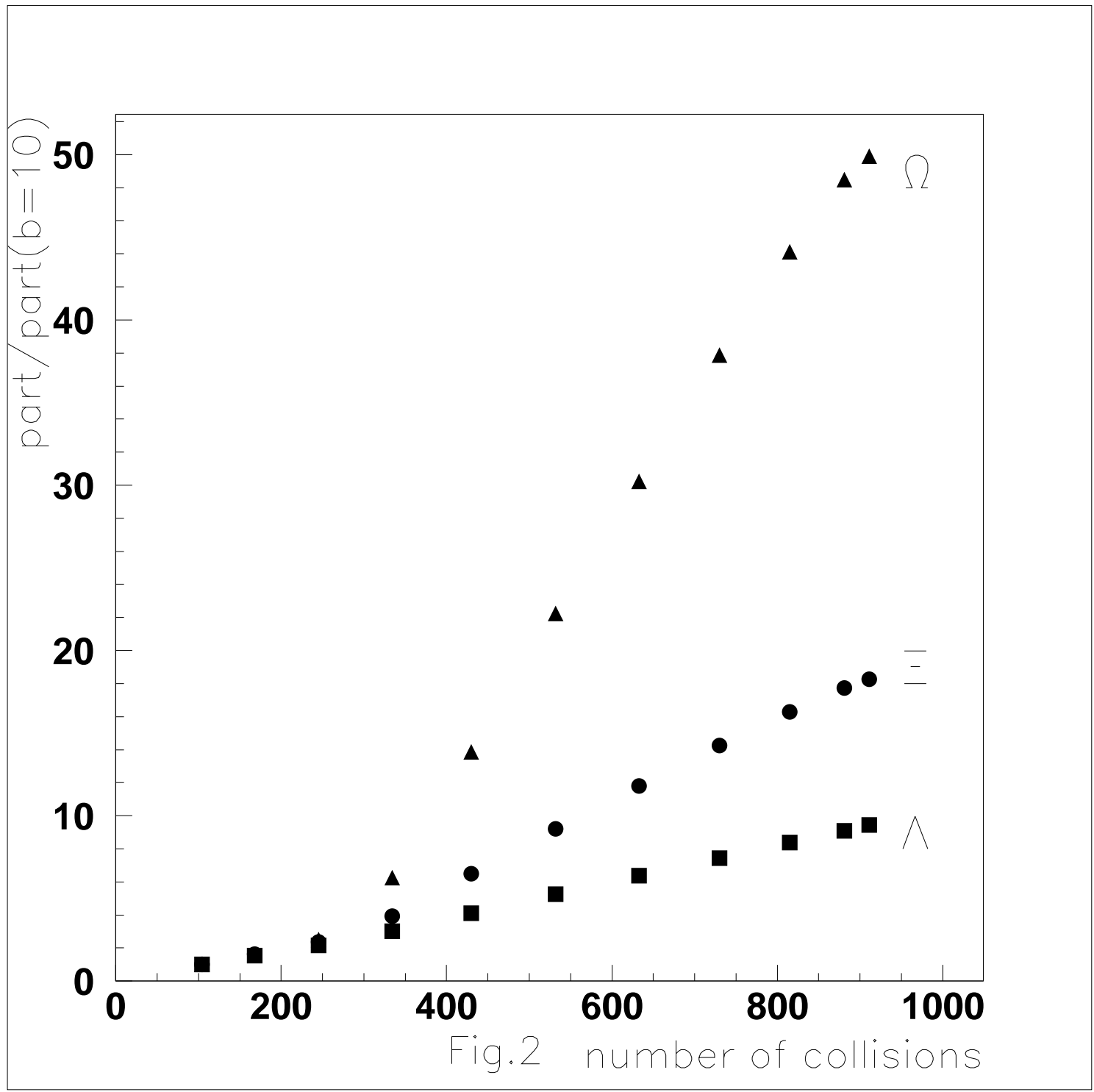
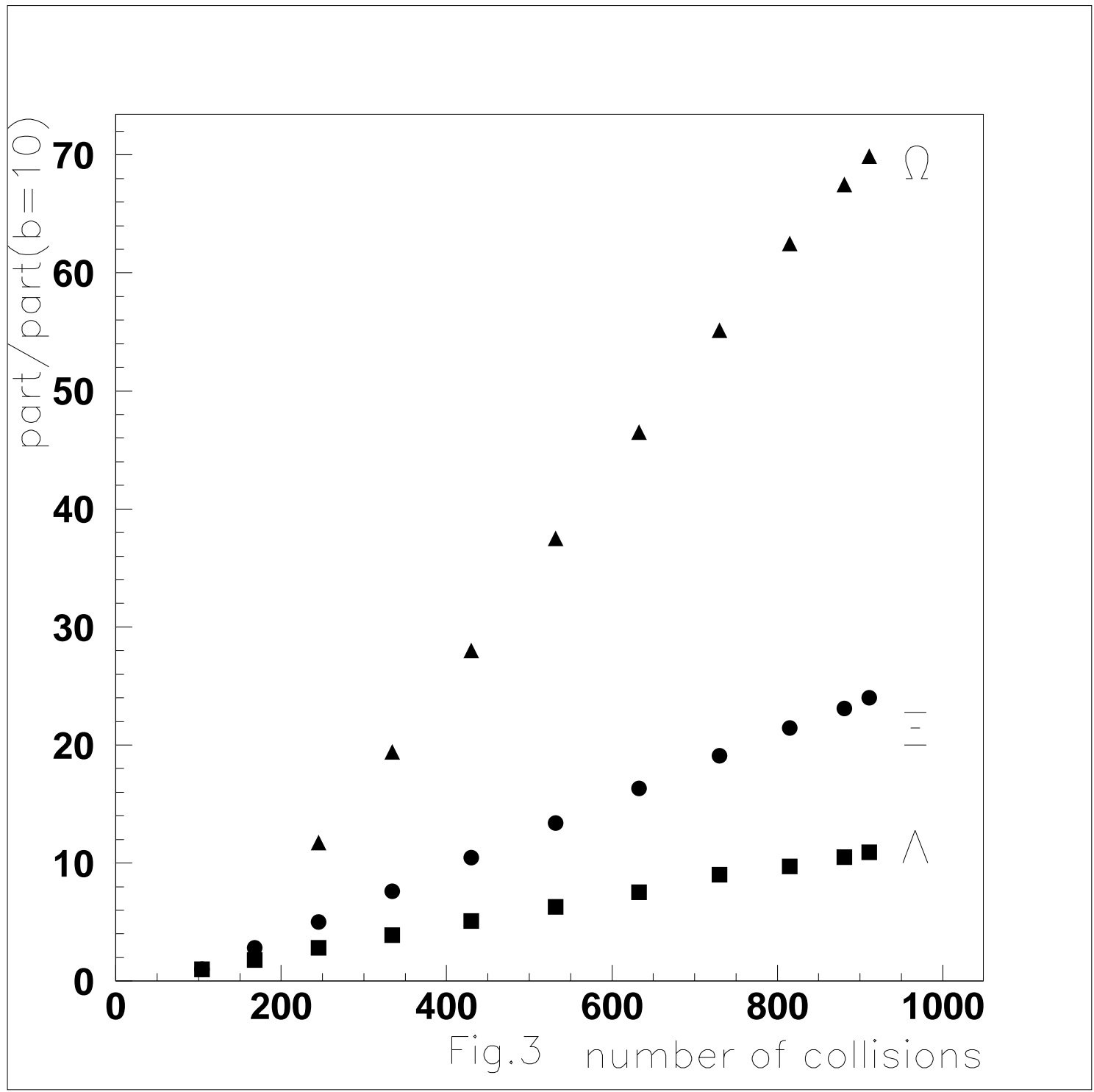
\end{document}